\documentclass[useAMS,usenatbib]{mn2e}

\usepackage{epsfig}
\usepackage{longtable}
\usepackage{times}

\usepackage{amsmath}
\usepackage{amssymb}

\def \xte {XTE\,J1739--302}
\def \igr {IGR\,J17544--2619}
\def \sax {SAX\,J1818.6--1703}

\def \inte {\textit{INTEGRAL}}

\def \degmark {^\circ}

\def \hcm {\hbox {\ifmmode $ atom cm$^{-2}\else atom cm$^{-2}$\fi}}

\def \apj {ApJ}

\def \apjl {ApJL}
\def \apjs {ApJS}
\def \aap {A\&A}

\def \mnras {MNRAS}

\def \aapr {A\&A Review}

\def \ima {\textsl{IMA}}

\newcommand{\be}{\begin{equation}}

\newcommand{\ee}{\end{equation}}

\title[$INTEGRAL$ study of temporal properties of flares in SFXTs]{$INTEGRAL$ study of temporal properties of bright flares in Supergiant Fast X-ray Transients }
\author[Sidoli et al.]{L.~Sidoli,$^{1}$\thanks{E-mail: sidoli@iasf-milano.inaf.it} A.~Paizis$^{1}$ and  K.~Postnov$^{2, 3}$\\
$^{1}$INAF, Istituto di Astrofisica Spaziale e Fisica Cosmica, Via E.\ Bassini 15,   I-20133 Milano,  Italy   \\
$^{2}$ Moscow Lomonosov State University, Faculty of Physics, Leninskie Gory 1, 117234, Moscow Russia \\
$^{3}$ Moscow Lomonosov State University, Sternberg Astronomical Institute, 117234, Moscow, Russia \\
}

\begin{document}

\date{Accepted 2016 January 26. Received 2016 January 25 ; in original form 2015 December 15 }

\pagerange{\pageref{firstpage}--\pageref{lastpage}} \pubyear{2016}

\maketitle

\label{firstpage}

\begin{abstract}
We have characterized the typical temporal behaviour of the bright  X--ray flares detected from the three 
Supergiant Fast X--ray Transients showing the most extreme transient behaviour (\xte, \igr, \sax).
We focus here on the cumulative distributions of the waiting-time (time interval between two consecutive X--ray flares),
and the duration of the hard X--ray activity (duration of the brightest phase of an SFXT outburst), 
as observed by \inte/IBIS in the energy band 17--50 keV. 
Adopting the cumulative distribution of waiting-times, it is possible to identify the typical timescale that clearly separates different 
outbursts, \text{each} composed by several single flares at $\sim$ks timescale.
This allowed us to measure the duration of the  brightest phase 
of the outbursts from these three targets, 
finding that they show heavy-tailed cumulative distributions.
We observe a correlation between the total energy emitted during SFXT outbursts and the time interval covered by the outbursts
(defined as the elapsed time between the first and the last flare belonging to the same outburst as observed by $INTEGRAL$).
We show that temporal properties of flares and outbursts of the sources, 
which share common properties regardless different orbital parameters, 
can be interpreted in the model of magnetized stellar winds with fractal structure from the OB-supergiant stars.
\end{abstract}


        %%%%%%%%%%%%%%%%%%%%%%%%%%%%%%%%%%%%%%%%%%%%%%%%%%%%%%%%%
        \section{Introduction\label{intro}}
        %%%%%%%%%%%%%%%%%%%%%%%%%%%%%%%%%%%%%%%%%%%%%%%%%%%%%%%%%

Supergiant Fast X--ray Transients (SFXTs; \citealt{Sguera2005}, \citealt{Negueruela2005a}) 
are a sub-class of high mass X--ray binaries (HMXBs) that was unveiled after the discovery 
by the \inte\ satellite of many short hard X--ray transients in the Galactic plane 
\cite[the first \inte\ source of this class was discovered by][]{Sunyaev2003}.

SFXTs host a neutron star (NS, hereafter) accreting from the wind
of either an O or B-type supergiant and display rare outbursts
punctuated by short ($\sim$1-2 ks) luminous flares reaching 10$^{36}$--10$^{37}$~erg~s$^{-1}$,
characterized by a duty cycle lower than a few \% 
(\citealt[][hereafter PS14]{Paizis2014}; \citealt{Lutovinov2013}, \citealt{Romano2014a}).
When observed at softer X--rays (1-10 keV) with more sensitive instruments, 
SFXTs are caught at X--ray luminosities below 10$^{34}$~erg~s$^{-1}$ most of the time  
(\citealt{Sidoli2008:sfxts_paperI}, \citealt{Romano2014b}, \citealt{Bozzo2015}).
In some members of the class, a luminosity as low as 10$^{32}$~erg~s$^{-1}$ (1-10 keV) has been
observed, leading to a very high observed dynamic range (ratio between luminosity during outburst and quiescence)
up to six orders of magnitude, in the most extreme case of the SFXT \igr\ \citep{zand2005, Romano2015}.

The mechanism responsible for the SFXT flaring behavior, associated with the accretion
onto the NS by a fraction of the donor wind, has been discussed by many authors.
Some of them explain the SFXT flares with the particular properties 
of the compact object  
(e.g. propeller effect or magnetic gating mechanism \citep{Grebenev2007,Bozzo2008}), 
others invoke a variety of orbital geometries together with the clumpy properties of the wind of the massive star
(\citealt{zand2005,Walter2007,Sidoli2007,Negueruela2008,Ducci2009}).
More recently, an alternative model has been proposed to explain bright flares in SFXTs  \citep{Shakura2014},  
based on the instability of the quasi-spherical shell of captured matter which accumulates 
above the magnetosphere of a slowly rotating NS at low accretion rates \citep{Shakura2012}.

After more than ten years from the discovery of SFXTs with \inte, 
the \inte/IBIS public archive is providing us with an observational data-set of X--ray flares detected from SFXTs 
that is large enough to enable a statistical investigation.
PS14 have exploited the \inte\ archive, 
investigating the cumulative distributions of the hard X--ray luminosity (17--100 keV) of SFXT flares,
finding  that they follow a power-law distribution, completely different with respect to the log-normal luminosity 
distribution shown by the emission from persistent ``classical'' HMXBs (e.g., Vela X--1).
These SFXT properties were succesfully interpreted by PS14 and \citet{Shakura2014} 
in terms of the model of unstable settling  accretion onto slowly rotating magnetized NS \citep{Shakura2012}.

Here we investigate other important properties of the same sample of SFXT flares reported by PS14: 
the waiting-time between two consecutive flares and the duration 
of the brightest phase of the outbursts, as observed by \inte, 
in order to get more insights into the nature of these transients.
We refer to PS14 for the summary of the SFXT properties adopted in the present paper.

%%%%%%%%%%%%%%%%%%%%%%%%%%%%%%%%%%%%%%%%%%%%%%%%%%%%%%%%%%%%%%%%%%%%
  	 \section{Data Analysis  \label{sec:data}}
%%%%%%%%%%%%%%%%%%%%%%%%%%%%%%%%%%%%%%%%%%%%%%%%%%%%%%%%%%%%%%%%%%%%

We focus here on the hard X--ray flares caught by \inte/IBIS observations covering about 9 years, from December 2002 to April 2012 (see PS14). 
In the present work we consider the three SFXTs  
which show the highest dynamic range (\xte, \igr\ and \sax) and are often referred to as ``prototypical SFXTs'' in the literature.

Data selection and analysis  have already been reported by PS14, 
adopting the data reduction procedure described 
in detail by \cite{Paizis2013}. 
We refer the reader to these papers for the technical details. 
In brief, the data-sets of detected X--ray flares consist of all IBIS pointing images (the so-called Science Windows, hereafter ScWs, with a typical
exposure time of $\sim$2~ks) where the sources were within $12\degmark$~from the image centre and found to be 
active (detection significance $>$5$\sigma$). Therefore, in this work we use the image deconvolution results ($\ima$~results) as reported by PS14.
As already discussed in PS14, since the typical time scale of flare duration is 
consistent with a ScW exposure time, we consider that a single IBIS/ISGRI detection on ScW level 
is representative of an SFXT flare.

The characterization of the luminosity distributions in hard X-rays of the  SFXT flares and their energy release  
have been already discussed  by PS14  and \citet{Shakura2014}, respectively. 

In the present work, we consider the same data-set of X--ray flares selected and reported by PS14 for the three
sources \xte, \igr\ and \sax, 
with the aim of further characterizing their temporal properties.

%%%%%%%%%%%%%%%%%%%%%%%%%%%%%%%%%%%%%%%%%%%%%%%%%%%%%%%%%%%%%%%%%%%%
  	 \section{Results \label{sec:res}}
%%%%%%%%%%%%%%%%%%%%%%%%%%%%%%%%%%%%%%%%%%%%%%%%%%%%%%%%%%%%%%%%%%%%

\begin{figure}
\centering
\begin{tabular}{cc}
\includegraphics[height=6.3cm, angle=0]{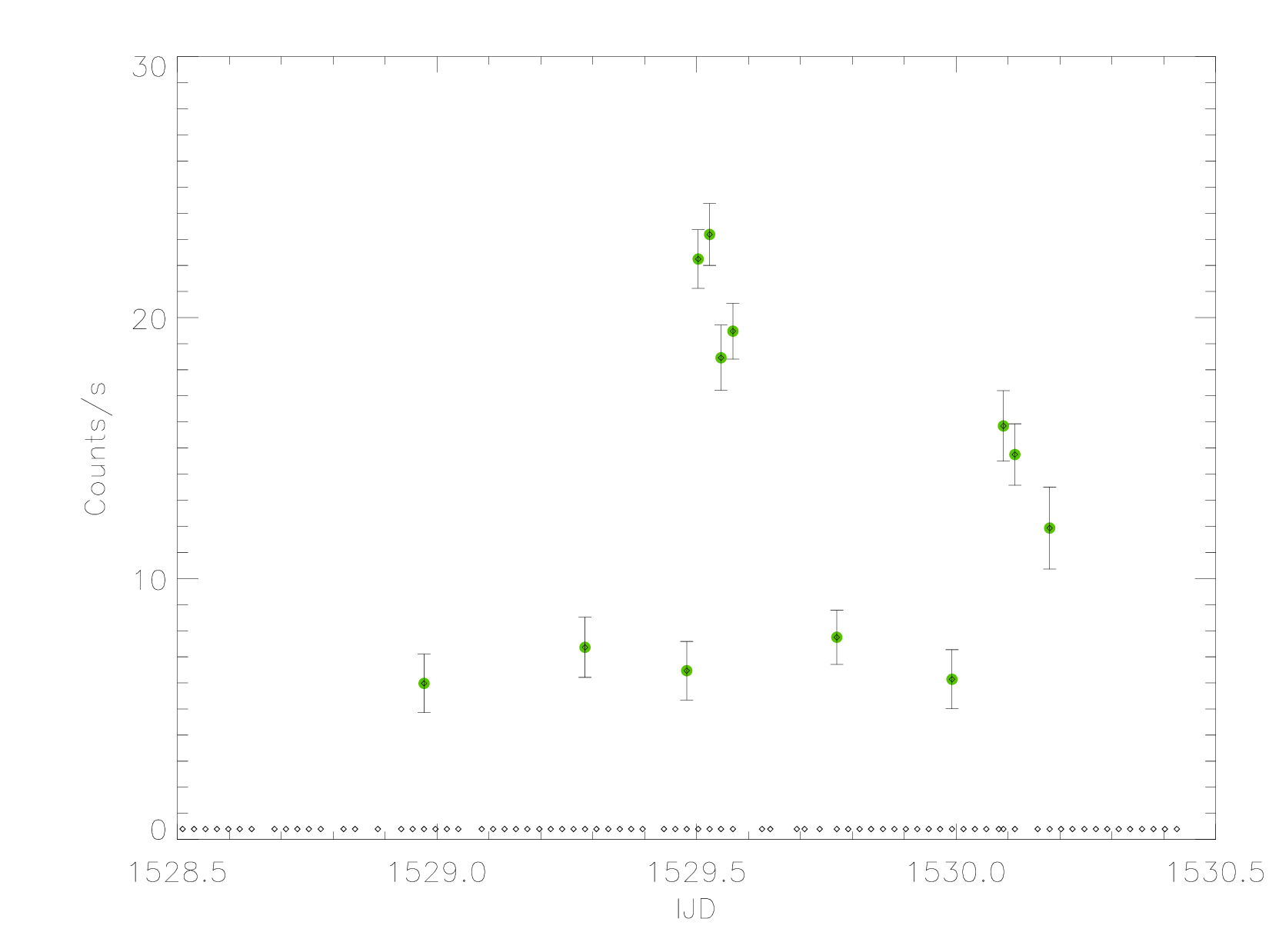} \\
\includegraphics[height=6.3cm, angle=0]{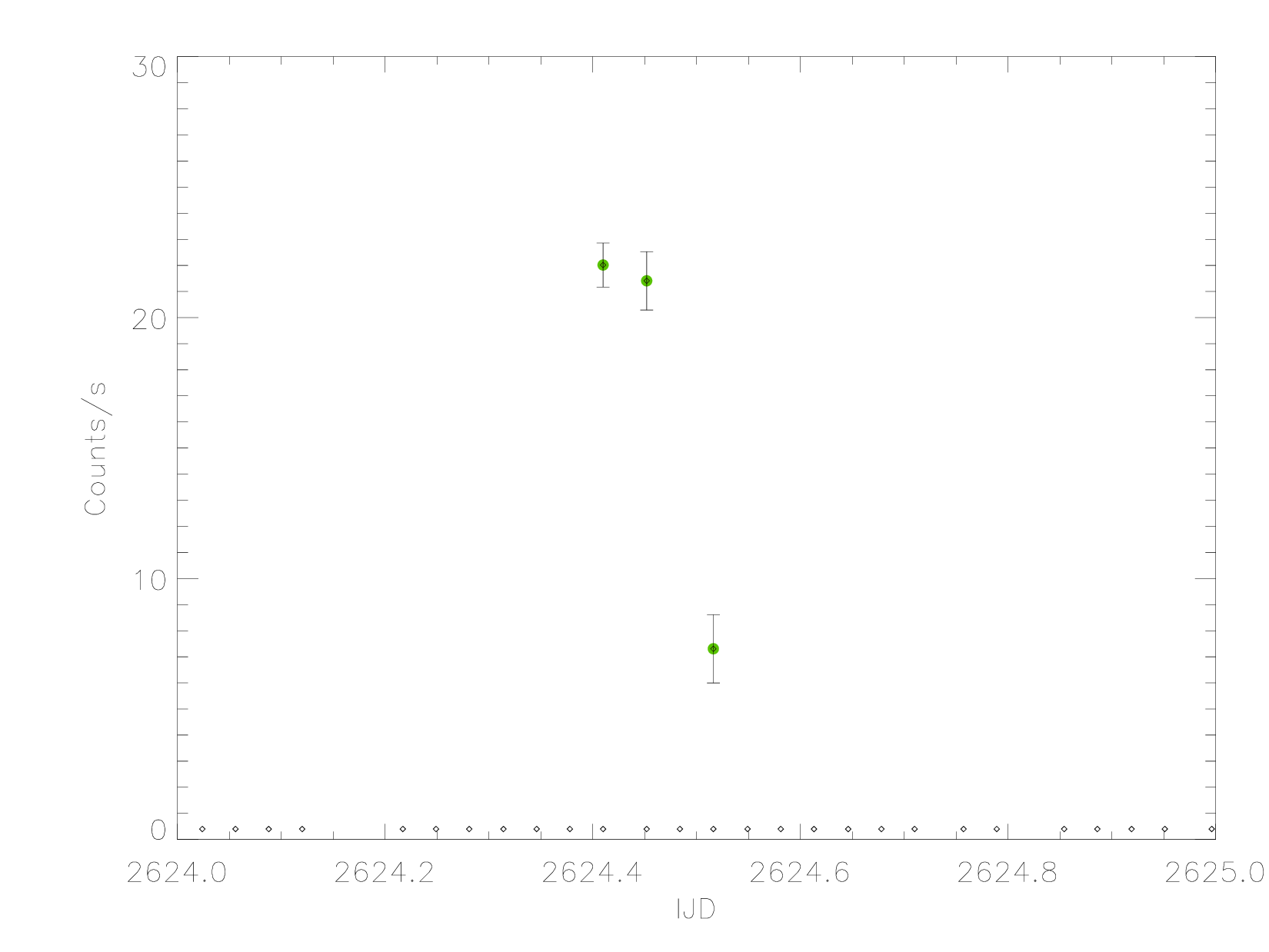}
\end{tabular}
\caption{Examples of SFXT light curves during outbursts (\xte), as observed by \inte\  (17--50 keV): a longer ({\em upper panel}) and a shorter outburst ({\em lower panel}). Time is in units of \inte\ day (IJD = MJD - 51544).  Diamonds along the x-axis mark the times of the \inte\ observations (ScWs) of the source field, while the solid green circle represents the detections, 
i.e. the X-ray flares (the  source count rate has been extracted over a ScW timescale, see text). The upper panel clearly shows the difference between two timescales defined in the text (the total duration ``D'' of the active phase of this outburst is 16~ks, while the elapsed time ``$\Delta$t'' between the last and the first flare of the same outburst is 10$^{5}$~s; see text for details). } 
\label{fig:lc_ex}
\end{figure}

\begin{figure}
\centering
\begin{tabular}{ccc}
\includegraphics[height=5.3cm, angle=0]{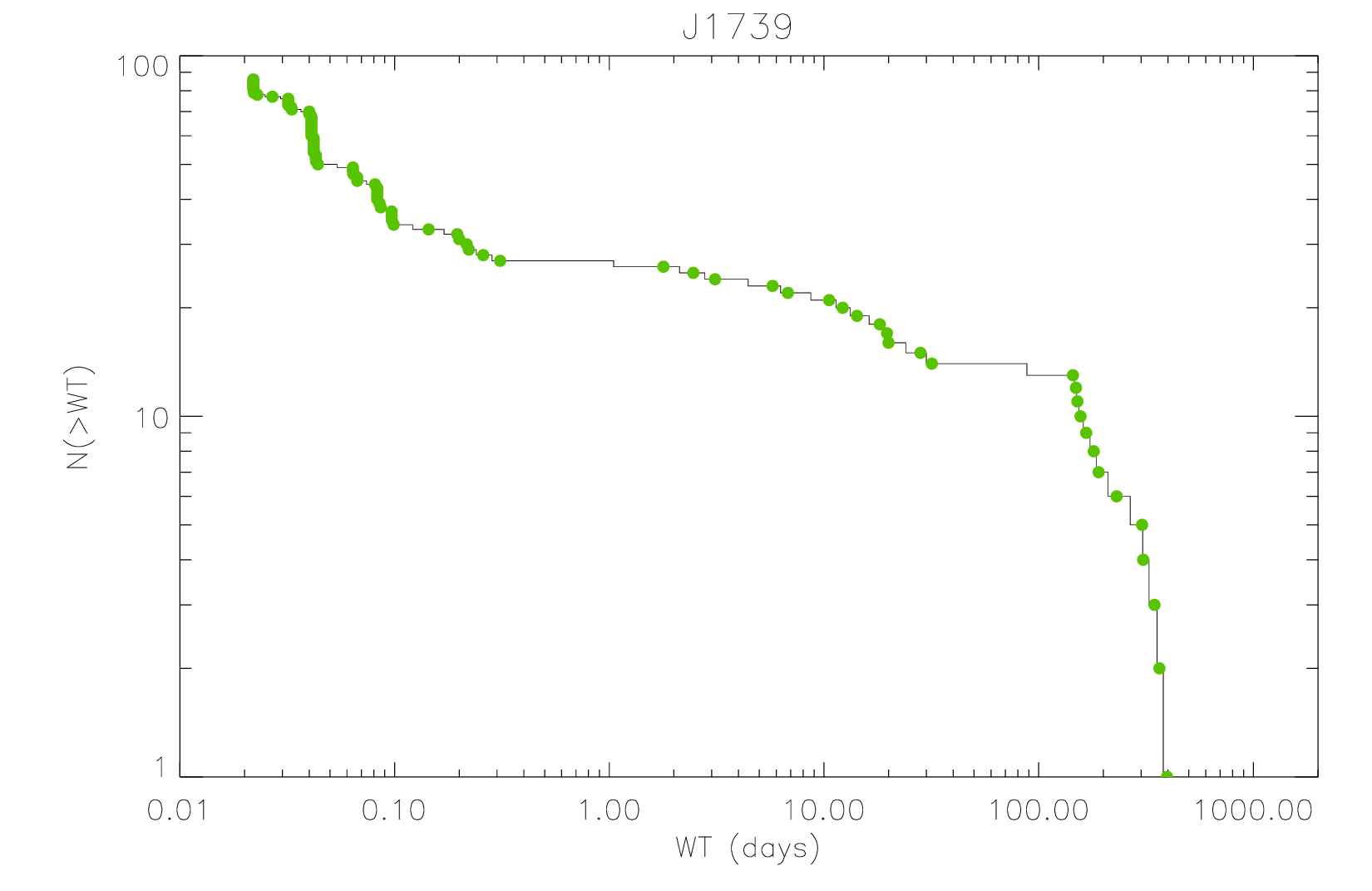} \\
\includegraphics[height=5.3cm, angle=0]{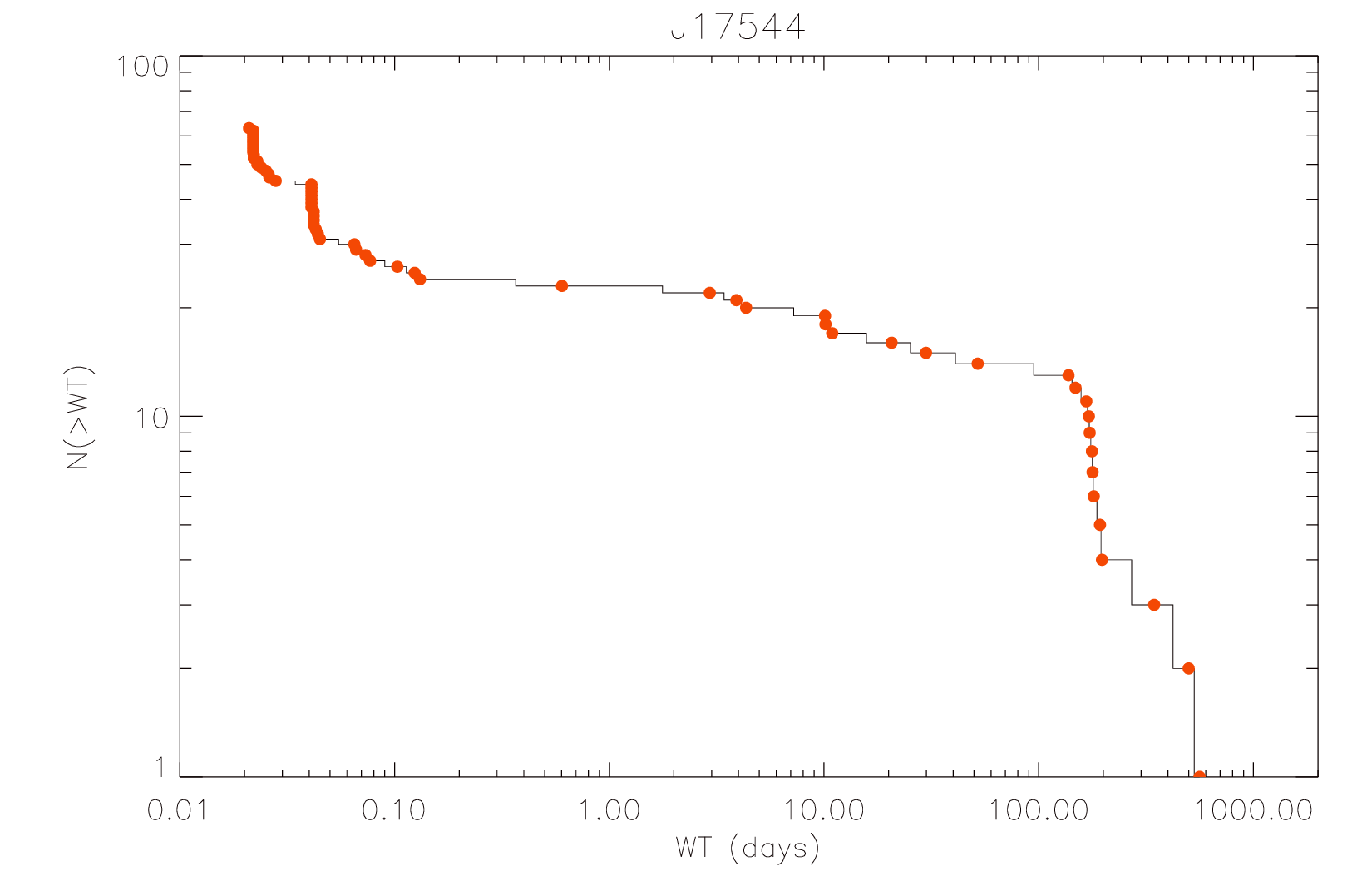} \\
\includegraphics[height=5.3cm, angle=0]{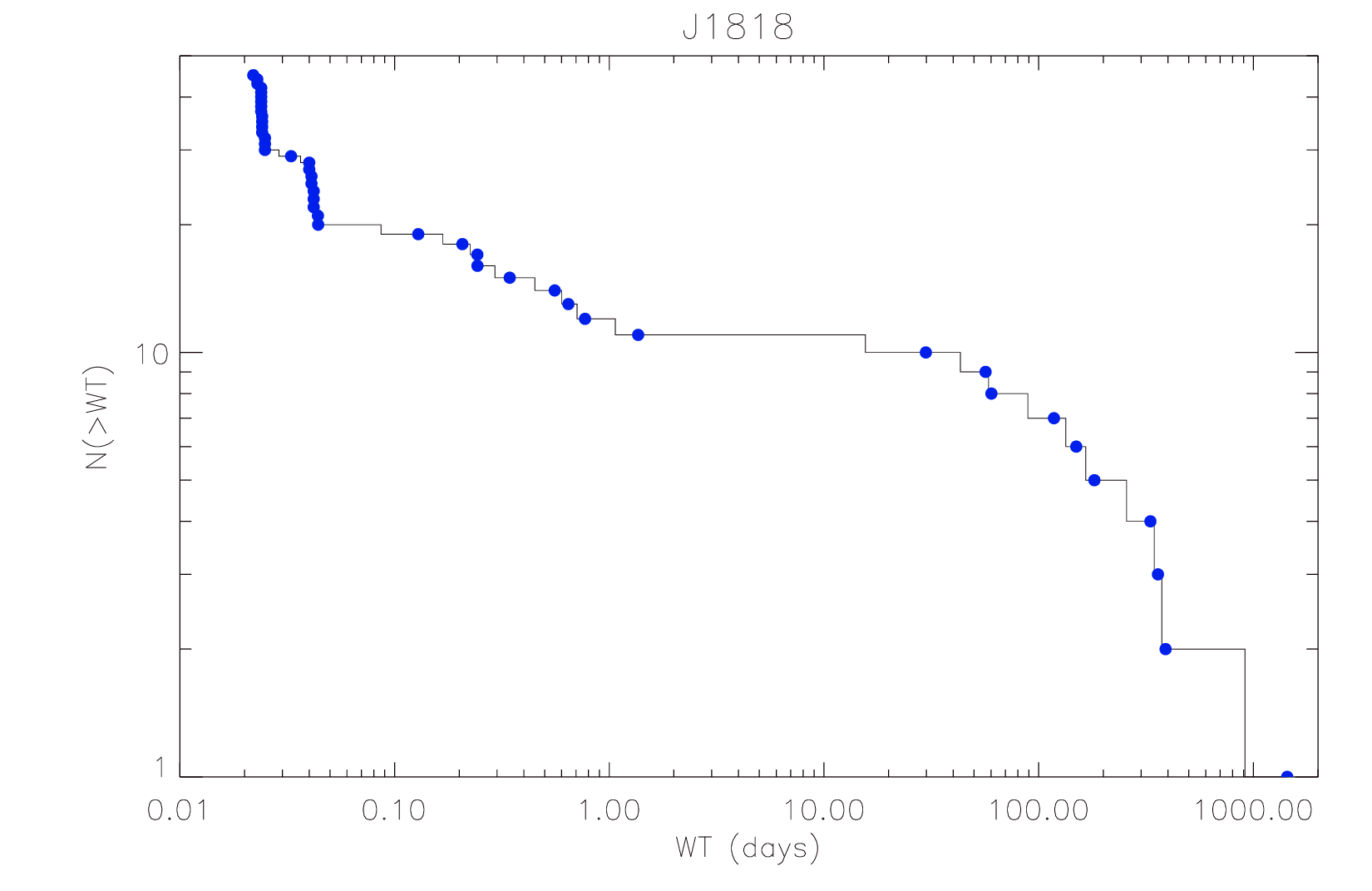}
\end{tabular}
\caption{Cumulative distributions of the waiting-times (WT) between two consecutive  
bright flares in the three SFXTs 
as observed by \inte/IBIS in about nine years of data (17--50\,keV).}
\label{fig:wt}
\end{figure}

We consider  the light curves of the X--ray flares (luminosity is in the energy band 17--50 keV) 
detected from the three SFXTs during \inte/IBIS observations spanning about 9 years.
Typical light curves during SFXT outbursts are shown in Fig.~\ref{fig:lc_ex} (where
each detection corresponds to a flare on ScW timescale). 
We define waiting-time (WT) between two consecutive flares, the difference 
between the start times of two subsequent ScWs where
a SFXT flare is detected. 
The shortest waiting-time corresponds to the duration of a ScW (in case of flares detected in adjacent ScWs). 
In Fig.~\ref{fig:wt} we show  the cumulative distribution of waiting-times between ``consecutive'' flares, for each of the three targets.

The highly populated vertical lines in Fig.~\ref{fig:wt},  corresponding to the shortest waiting-time (below 0.1 days), 
are produced by flares detected in adjacent ScWs.
At the opposite side of the x-axis,  waiting-times larger than 100~days might be significantly affected by the
data gaps between two satellite visibility windows of the source sky position, as suggested by the steepening of
the waiting-time distribution above 100~days, particularly evident in  \xte\ and  \igr. 
Other biases due to gaps in the data are unlikely to severily affect the shape of the 
true distribution at intermediate waiting-times (between a few days and 100~days), given the low duty cycles of these three SFXTs.

An interesting feature present in all three waiting-time distributions is 
a plateau just above $\sim$1~day (i.e. missing waiting-times in that temporal range). 
This strongly suggests that a waiting-time of $\sim$1~day 
(1.4~day, to be precise, in order to include the last point of the plateau in \sax)
can be adopted as a timescale to separate the flaring 
activity belonging to two  different and subsequent outbursts
(each composed by a cluster of many flares). 

This makes it  possible to derive the time duration of single outbursts and to study their 
statistical properties (e.g., cumulative distribution) and correlations. 
In order to determine the total duration of a bright X--ray flaring activity episode (a single outburst) 
as observed by \inte/IBIS, we adopt the following procedure: 
for each SFXT, we start from the second X--ray flare detected in the \inte/IBIS light curve and calculate its 
waiting-time with respect to the first flare. 
If this waiting-time is lower than 1~day, then the flare is assumed to belong to the
same bright outburst phase as the previous flare. 
Performing this comparison with the following flares, at one point there will be a flare for which the waiting-time with respect
to the immediatly previous flare is greater than 1 day. At this point the procedure stops and a list of clusters of flares (i.e. outburst) 
is produced for the three SFXTs.

This automatic process has led to the empirical determination of the duration of the SFXTs outbursts, as observed by \inte.
To do this, we defined two different temporal quantities: 
the {\em total duration} (D) of the brightest phase of an outburst and the {\em elapsed time interval} ($\Delta $t) between the first and the last
flare belonging to an outburst  (see Fig.~\ref{fig:lc_ex}, upper panel and caption, for an example).
More in detail: for each source, for each single outburst, we calculated the total duration (D) simply by adding together
the durations, d$_i$, of single flares belonging to the same outburst (D=$\sum _{i=1}^N d_i$, where N is the number 
of flares in a single outburst). 
Since the  flares composing an outburst can be either detected in adjacent pointings or not, 
D can be interpreted as the integrated time each SFXT spends in its brightest state during a single outburst as observed by IBIS.
Durations D lower than 2-3~ks imply that only one isolated flare is caught by \inte\ during a supposed (although unobserved) longer outburst. 
Very likely, these single flares represent the brightest activity of outburts with average fainter luminosity, 
where \inte\ detects only the brightest part. 

Another temporal quantity that can be calculated from the light curves of X--ray flares, for each outburst, is
the elapsed time interval,  $\Delta $t (where $\Delta $t=tstop$_{N}$-tstart$_{1}$) between the first (``1'') and the last (``N'') 
flare belonging to
the same outburst (as previously defined).

%%%%%%%%%%%%%%%%%%%%%%%%%%%%%%%%%%%%%%%%%%%%%%%%%%%%%%%% 
%%%%%%%%%%%%%%%%%%%%%%%%%%%%%%%%%%%%%%%%%%%%%%%%%%%%%%%%

\subsection{Cumulative distributions}

\begin{figure}
\centering
\begin{tabular}{ccc}
\includegraphics[height=5.3cm, angle=0]{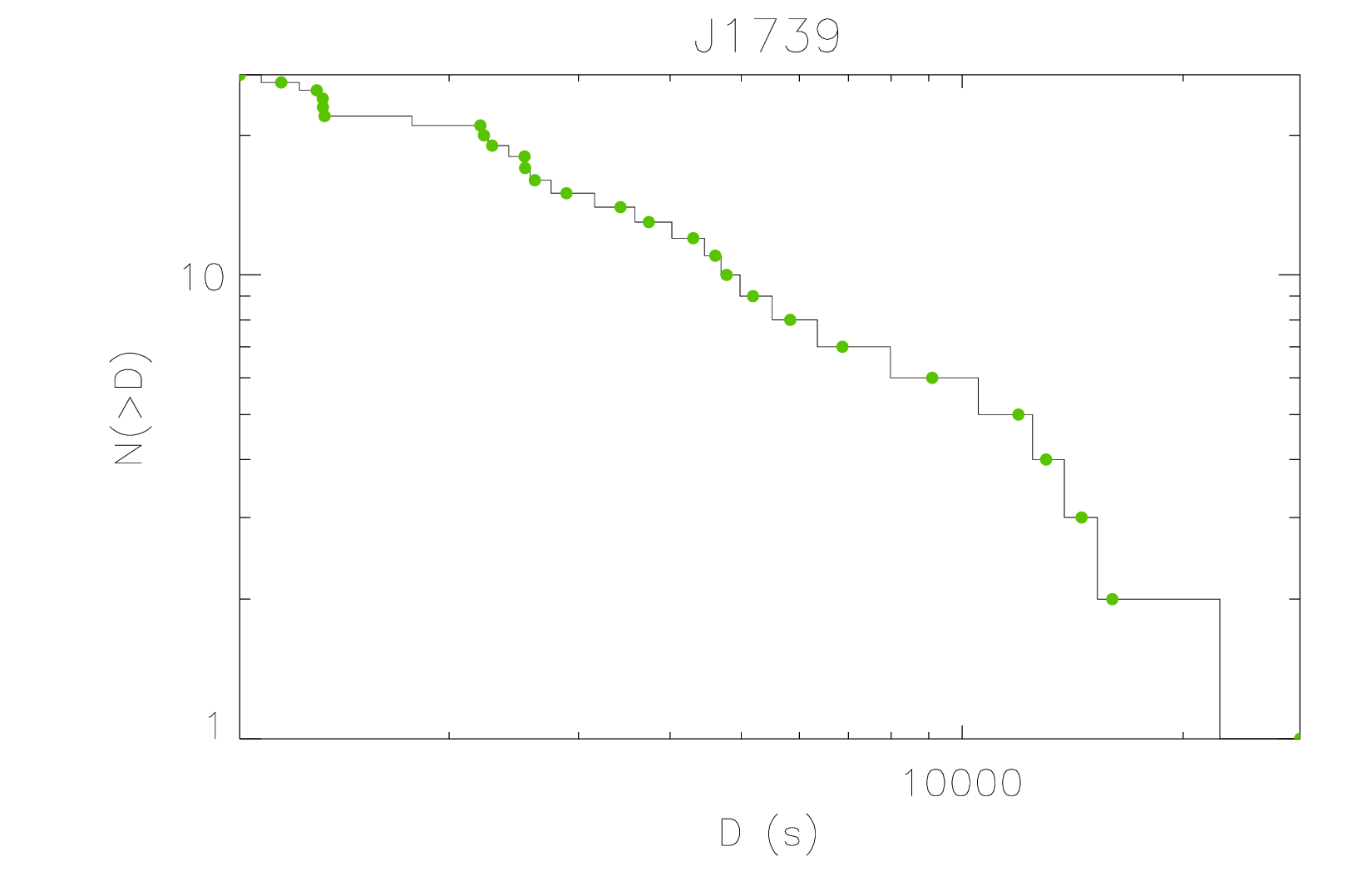} \\
\includegraphics[height=5.3cm, angle=0]{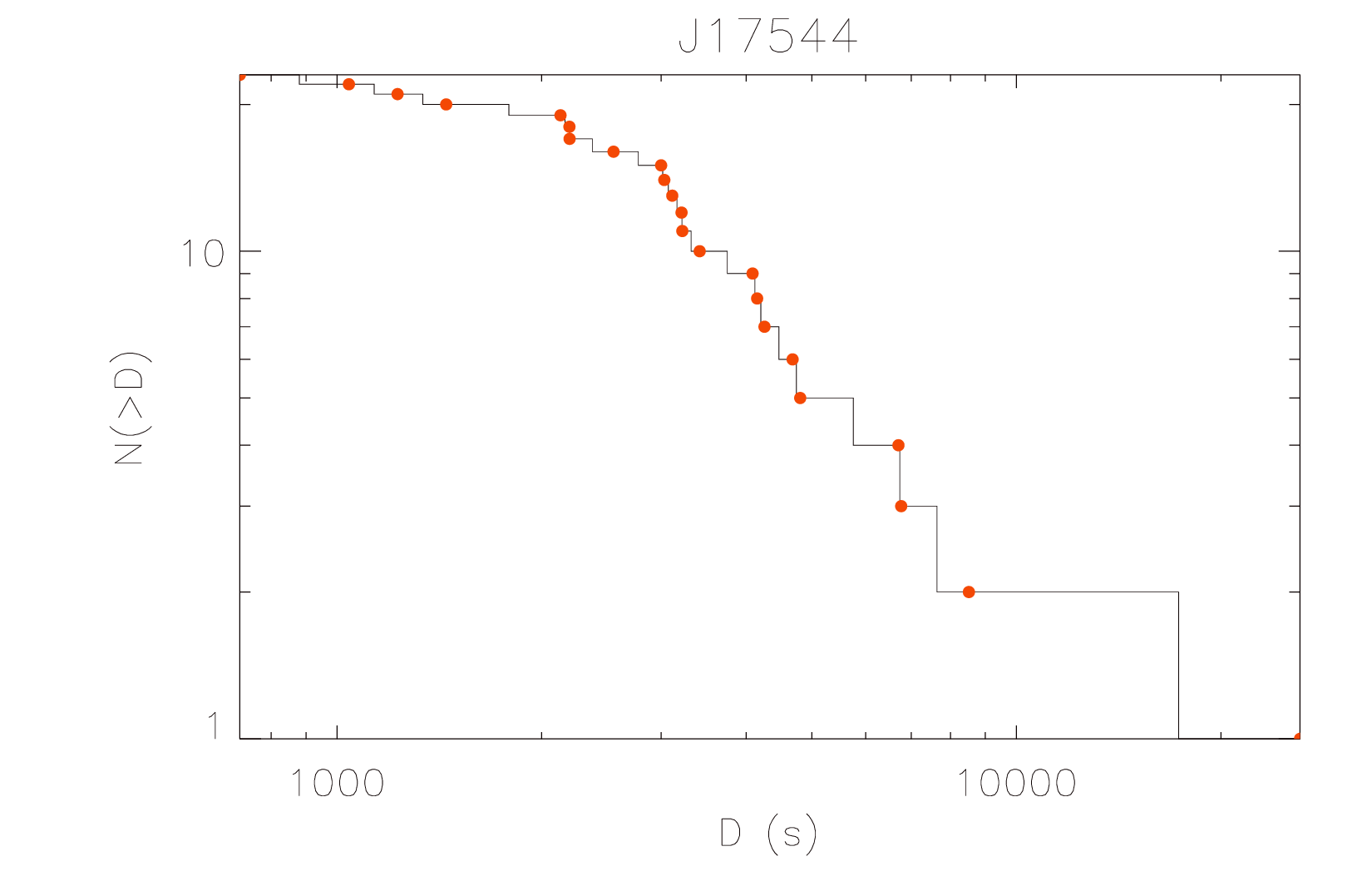} \\
\includegraphics[height=5.3cm, angle=0]{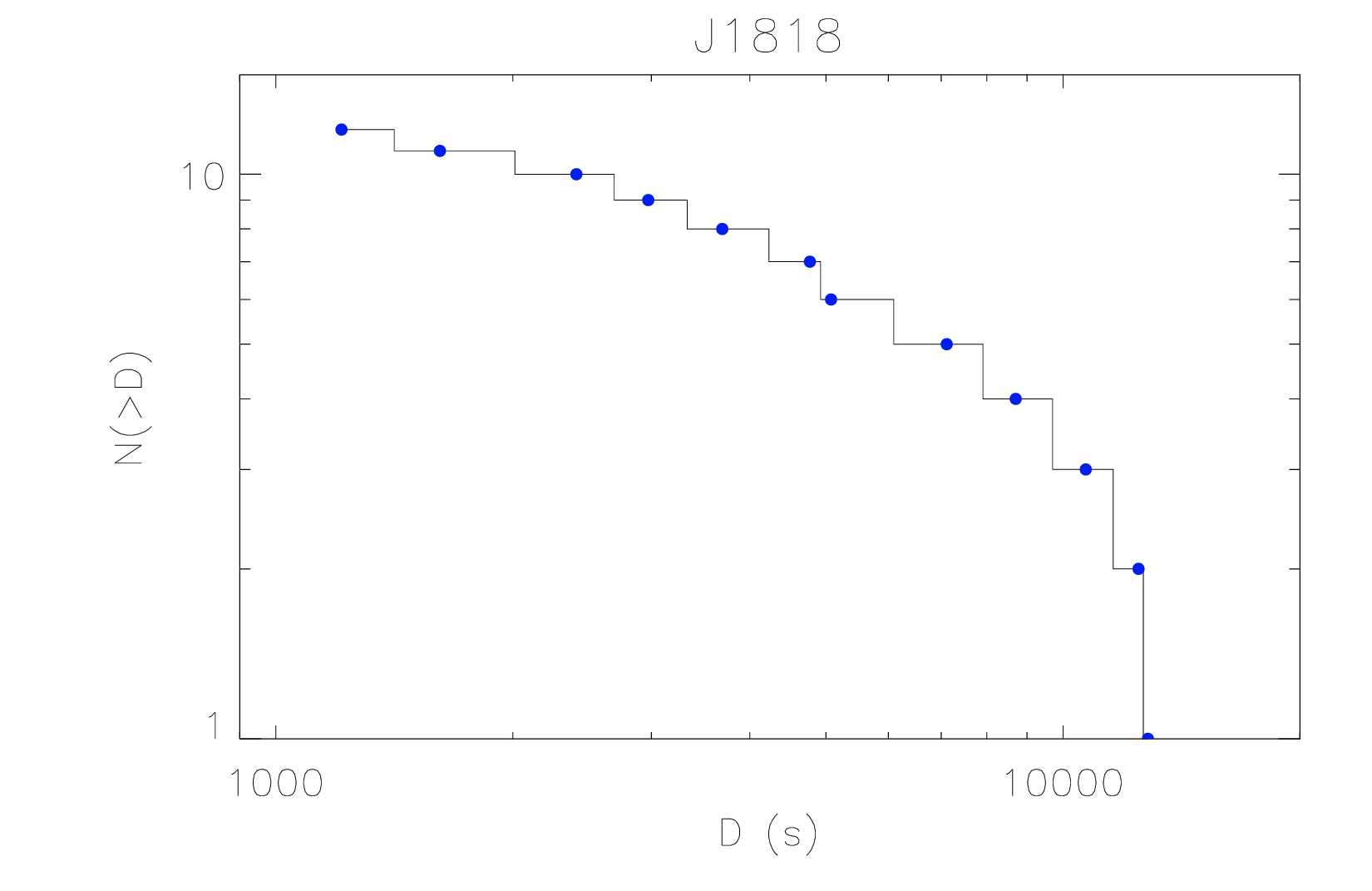}
\end{tabular}
\caption{Cumulative distributions of the duration (D)
of the outbursts  in the three SFXTs, as observed by \inte/IBIS (17--50\,keV).}
\label{fig:dur}
\end{figure}

The cumulative distributions of the durations D of the brightest phase in each single outburst
for the 3 SFXTs considered here are shown in Fig.~\ref{fig:dur}.
In particular, an evident 
flattening at short durations is present in the \igr\ duration distribution, due to single flares.
While the cumulative distribution of the outburst duration in \sax\ appears to be exponential, in the case of 
\igr\ and \xte, if we exclude the flatter region at low duration composed by single flares, they are power-law-like. 
Adopting a maximum-likelihood estimation  of the power-law slope (PS14) from a 
subsample of data points above a truncation point of 2~ks and 3~ks (respectively for \xte\ and \igr), 
we obtained a power-law slope, $\beta$, of 1.6$\pm${0.6} able to adequately describe 
the cumulative distribution outburst duration in \igr, while $\beta$=0.5$\pm{0.2}$ in \xte.
We note that in all three SFXTs, the cumulative distribution of the 
total durations of the outburst phases (in their brightest phase, the one that IBIS is able to detect) 
display a cut-off above 10~ks. In particular, the maximum outburst durations observed in our data-set 
 are 13~ks in \sax, 26~ks in \igr~and 29~ks in \xte.

\begin{figure}
\centering
\begin{tabular}{ccc}
\includegraphics[height=5.3cm, angle=0]{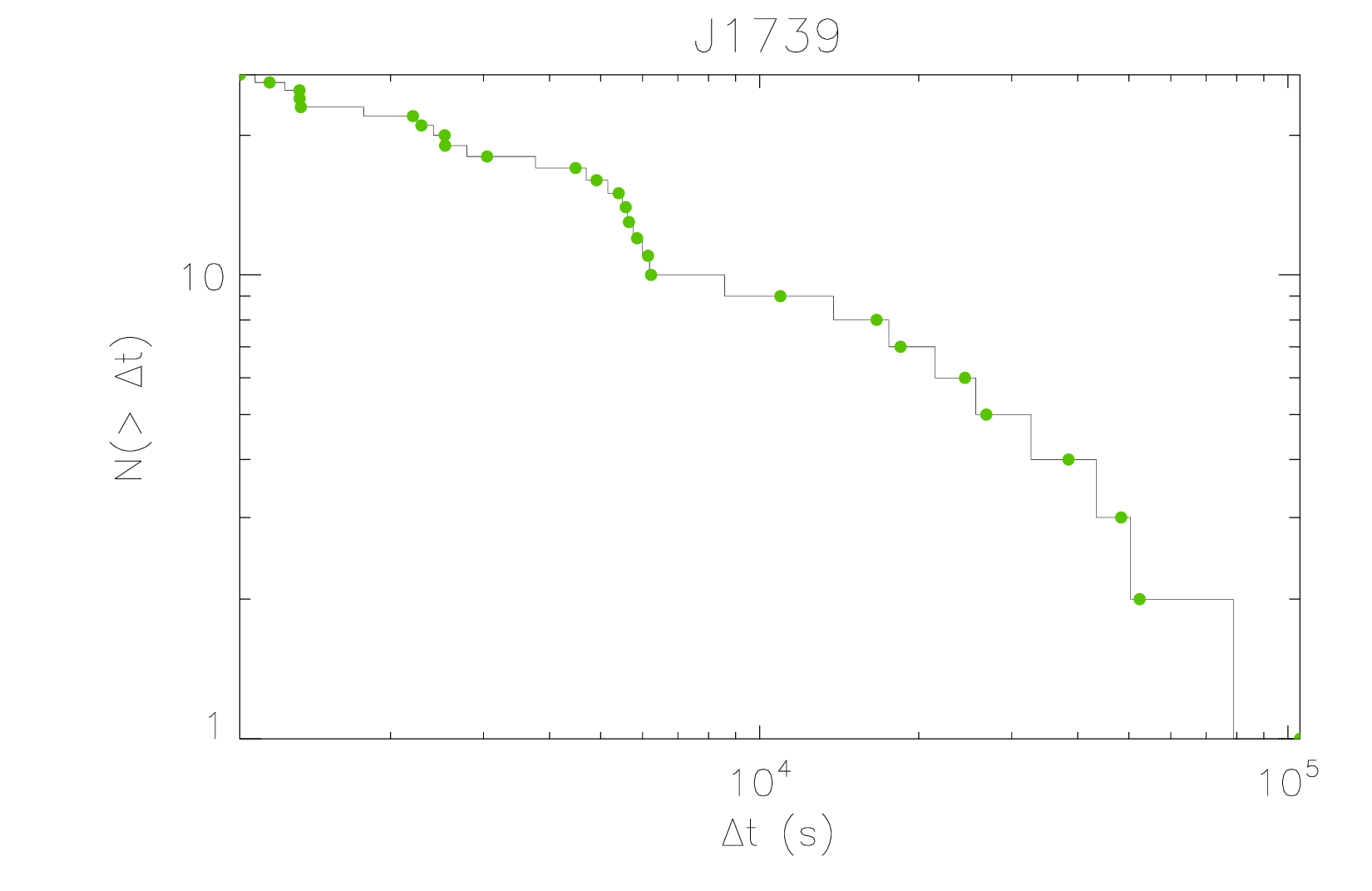} \\
\includegraphics[height=5.3cm, angle=0]{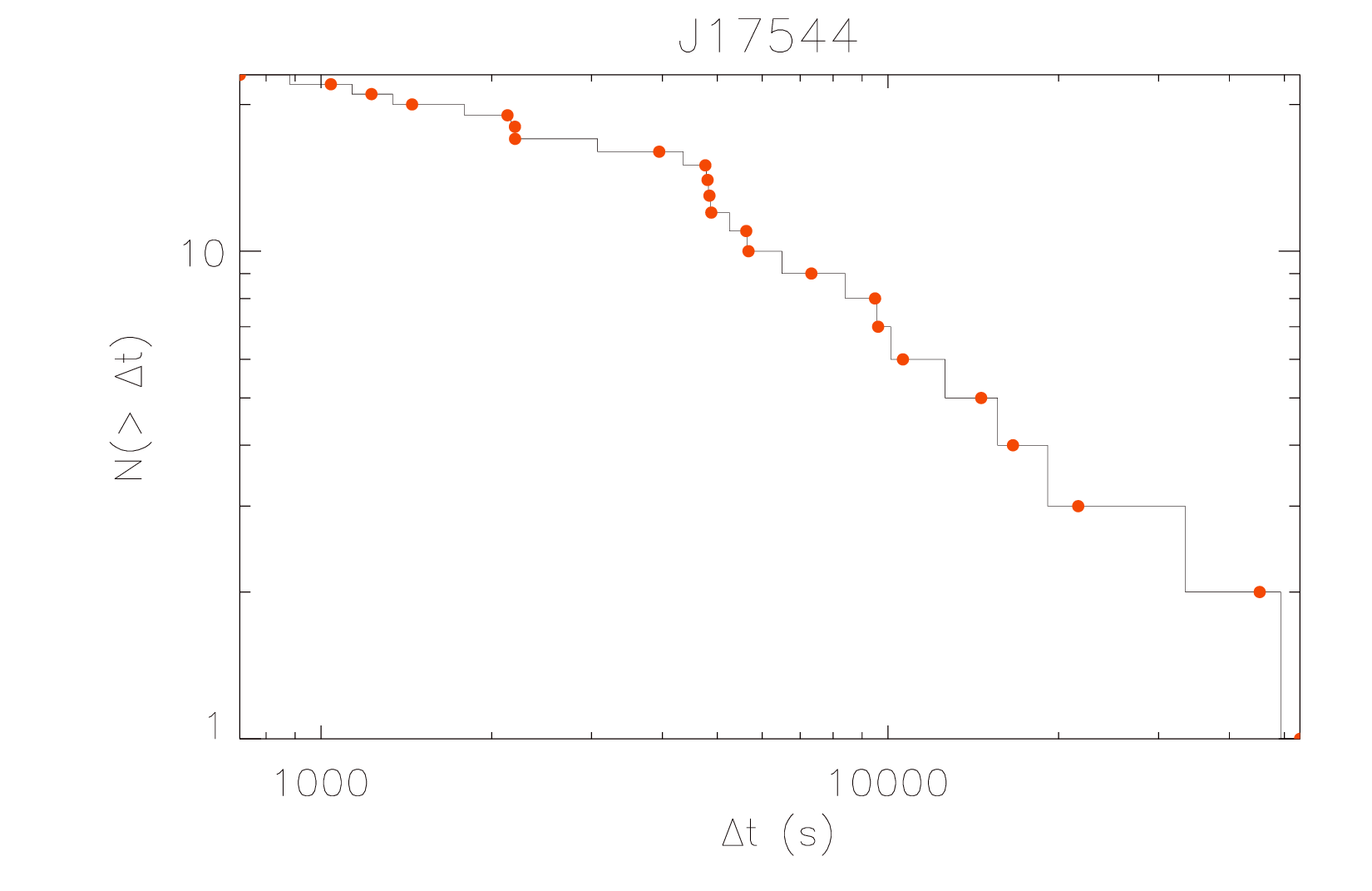} \\
\includegraphics[height=5.3cm, angle=0]{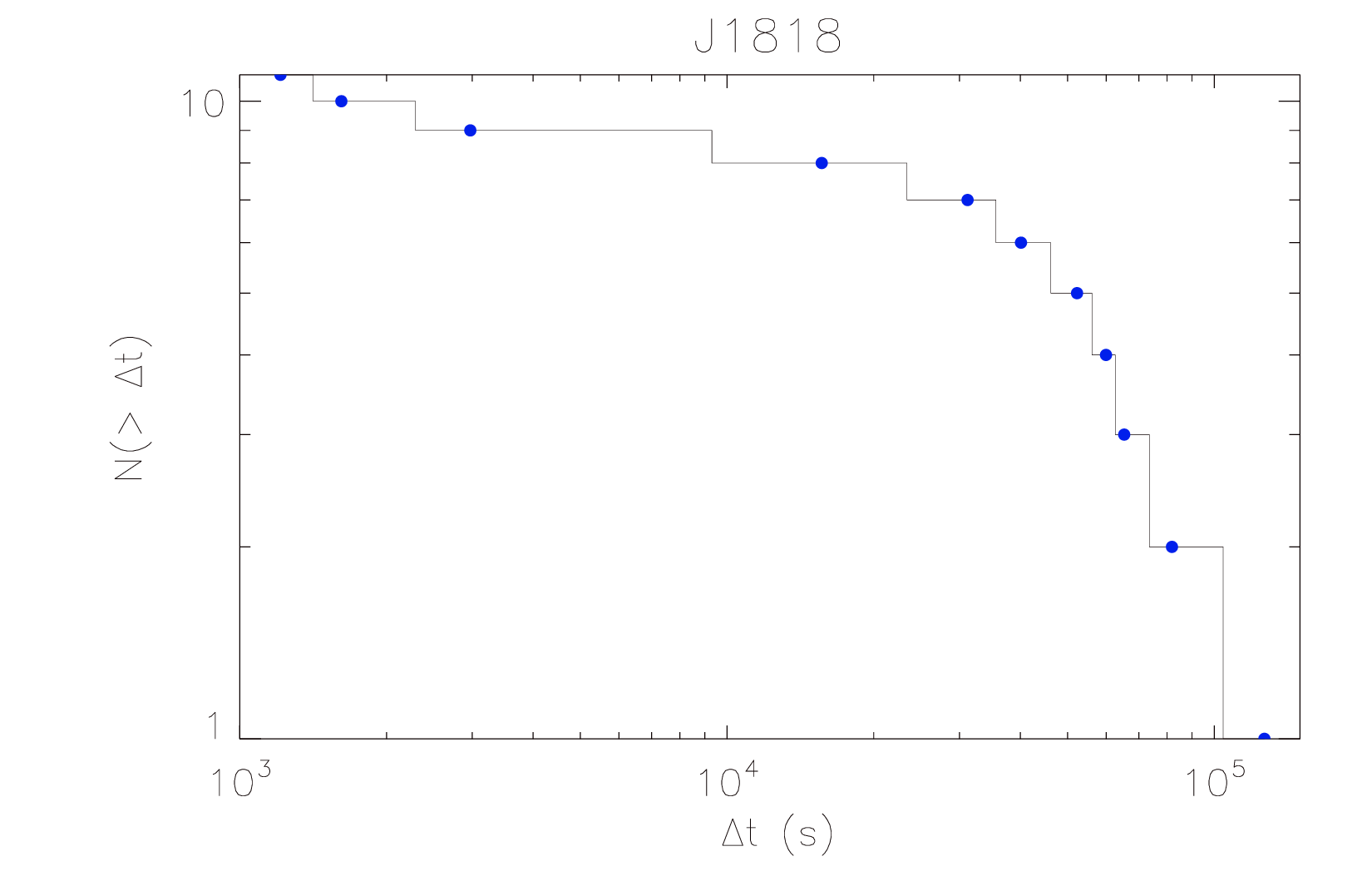}
\end{tabular}
\caption{Cumulative distributions of elapsed times ($\Delta$t) 
in the three SFXTs, as observed by \inte/IBIS (17--50\,keV).}
\label{fig:deltat}
\end{figure}

The distributions of the  elapsed times, $\Delta $t, for the three SFXTs are shown in Fig.~\ref{fig:deltat}.
Note that in case of outbursts made of a single isolated flare, again, $\Delta $t is equal to the duration 
of the single flare. This causes the flattening of the distributions at low  $\Delta $t around 1-2~ks (and below).
Above this timescale, the distribution of $\Delta $t in \xte\ shows a steepening (higher frequency of flares) 
around 5-6~ks, with a few outbursts composed by only 2-3 flares, 
while above 10~ks the cumulative distribution is power-law-like. 
In \igr\ a power-law-like distribution of $\Delta $t is present above around 5-6~ks, while in 
\sax\ a roll-over above about 30-40~ks emerges.

%%%%%%%%%%%%%%%%%%%%%%%%%%%%%%%%%%%%%%%%%%%%%%%%%%%%%%%%
%%%%%%%%%%%%%%%%%%%%%%%%%%%%%%%%%%%%%%%%%%%%%%%%%%%%%%%%

%%%%%%%%%%%%%%%%%%%%%%%%%%%%%%%%%%%%%%%%%%%%%%%%%%%%%%%%
%%%%%%%%%%%%%%%%%%%%%%%%%%%%%%%%%%%%%%%%%%%%%%%%%%%%%%%%

%%%%%%%%%%%%%%%%%%%%%%%%%%%%%%%%%%%%%%%%%%%%%%%%%%%%%%%%
%%%%%%%%%%%%%%%%%%%%%%%%%%%%%%%%%%%%%%%%%%%%%%%%%%%%%%%%

\begin{figure}
\centering
\begin{tabular}{cc}
\includegraphics[height=5.3cm, angle=0]{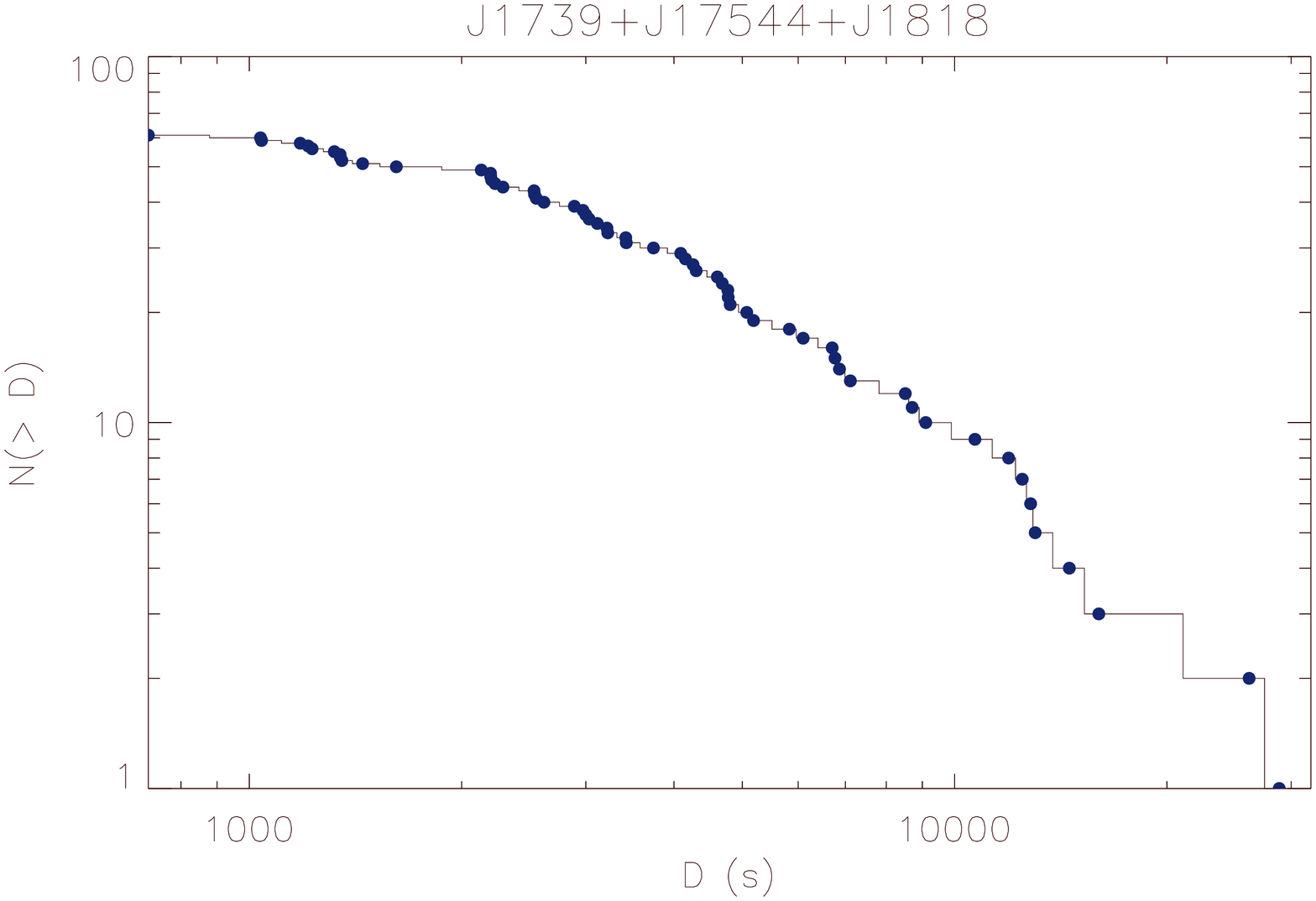} \\
\includegraphics[height=5.3cm, angle=0]{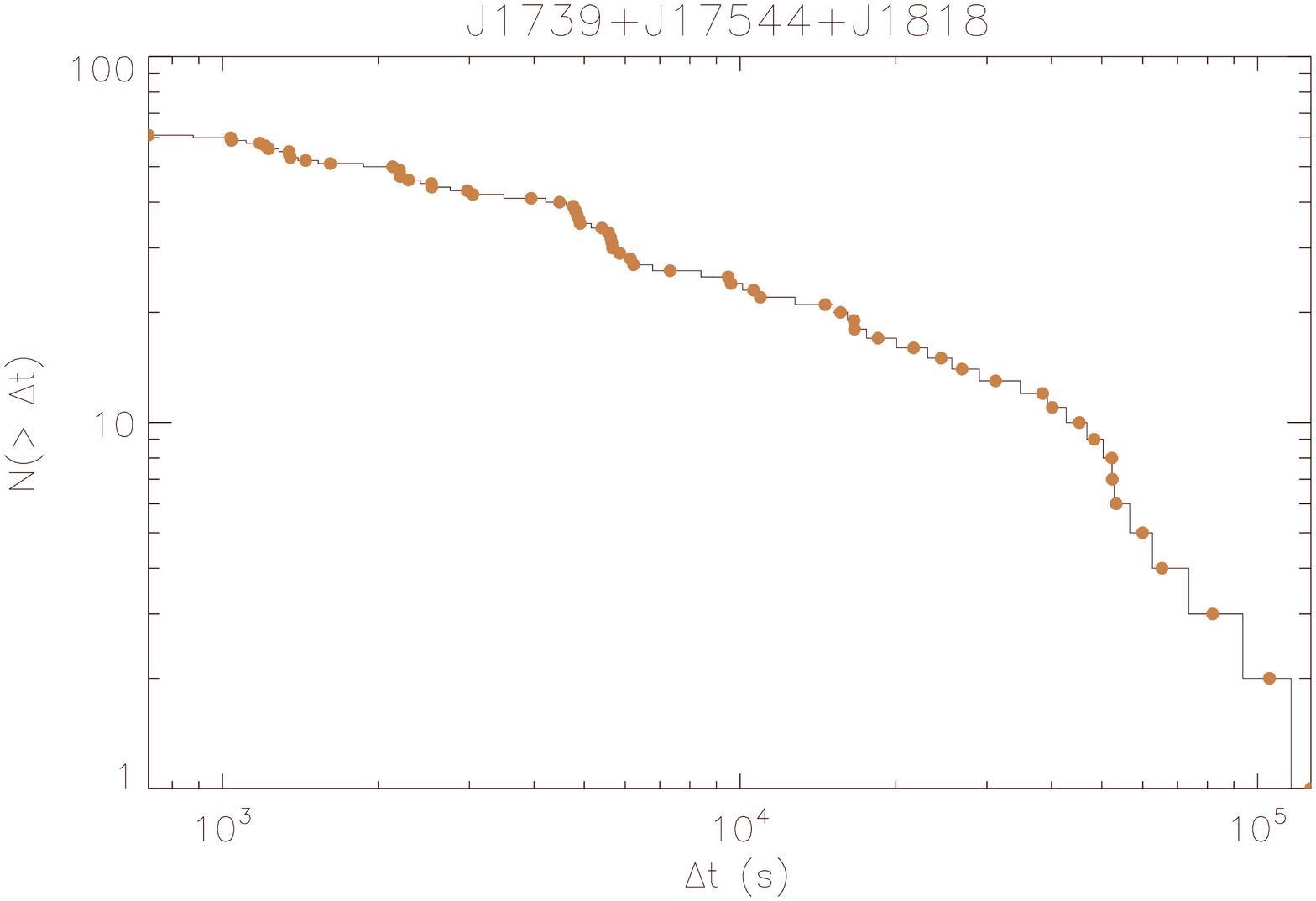}
\end{tabular}
\caption{Cumulative distributions of the durations (D$_{all}$, {\em upper panel}) 
and elapsed times ($\Delta$t$_{all}$, {\em lower panel}) obtained from combining
all outbursts observed from all the three SFXTs analysed here.}
\label{fig:all}
\end{figure}

In Fig.~\ref{fig:all},  we show the cumulative distribution of
the total durations (D$_{all}$, upper panel) and elapsed times ($\Delta$t$_{all}$, lower panel), 
respectively, taken from combining 
timescales characterizing the outbursts from all three SFXTs.
It is remarkable that a power-law distribution is able to adequately describe the global 
behaviour of the three prototypical
SFXTs, with slopes, $\beta$, of 0.9$\pm{0.3}$ and 0.3$\pm{0.1}$ for D$_{all}$ (above $\sim$2~ks) 
and $\Delta$t$_{all}$ (above $\sim$6~ks), respectively.
A roll-over is present at higher timescales (above $\sim$10~ks for the overall 
durations D$_{\rm all}$, while above $\sim$30-40~ks in the distribution of all elapsed times $\Delta$t$_{\rm all}$).
The overall regular shape of the two plots is suggestive of the fact that we
are seeing the same phenomenology at work in the three sources, as if to say
that by merging the results from the three prototypical SFXTs, we are
observing a single SFXT with a broader coverage, where almost no timescale is
missed due to observability gaps.

%%%%%%%%%%%%%%%%%%%%%%%%%%%%%%%%%%%%%%%%%%%%%%%%%%%%%%%%
%%%%%%%%%%%%%%%%%%%%%%%%%%%%%%%%%%%%%%%%%%%%%%%%%%%%%%%%

\subsection{The energy released in the outburst versus its duration}
 
The procedure of separation of individual outburst (a collection of physically connected flares) 
from the $INTEGRAL$ data outlined above enables us to study their different statistical properties. 
In Fig.~\ref{lsfig:fit} we plot the total energy released in the j-th outburst, $\Delta E_j$, 
as a function of its total duration, $\Delta T_j$. 
In Figs.~\ref{lsfig:fit} the total energy is reported in units of ergs (left panels) 
and of total IBIS/ISGRI counts (17--50~keV; right panels),
to clearly show that the main contribution to the uncertainty is the source distance 
(see PS14 for details on conversion factors from counts to physical units).

Here
$\Delta E_j = \sum_{i=1}^{N_j} \delta t_i f_i$ which is the total energy
released during the j-th outburst, $\delta t_i$ is the
duration of i-th flare in the outburst 
(which was taken to be the duration of an $INTEGRAL$
ScW) and $f_i$ is the mean luminosity (erg~s$^{-1}$) of a single flare (using a ``bin time'' of one ``ScW'').

In this plot, $\Delta T_j$ is the outburst duration (in the sense of
the ``elapsed time'' interval 
$\Delta$t shown in Fig.~\ref{fig:deltat}). 
It should be taken into account 
that the
shortest $\Delta T_j$ (2000 seconds and shorter) are actually the durations
of a single ScW, when the ``outburst'' detected by $INTEGRAL$ is composed by a single flare.

\begin{figure*}
\centering
\begin{tabular}{cc}
\includegraphics[height=11.cm, angle=0]{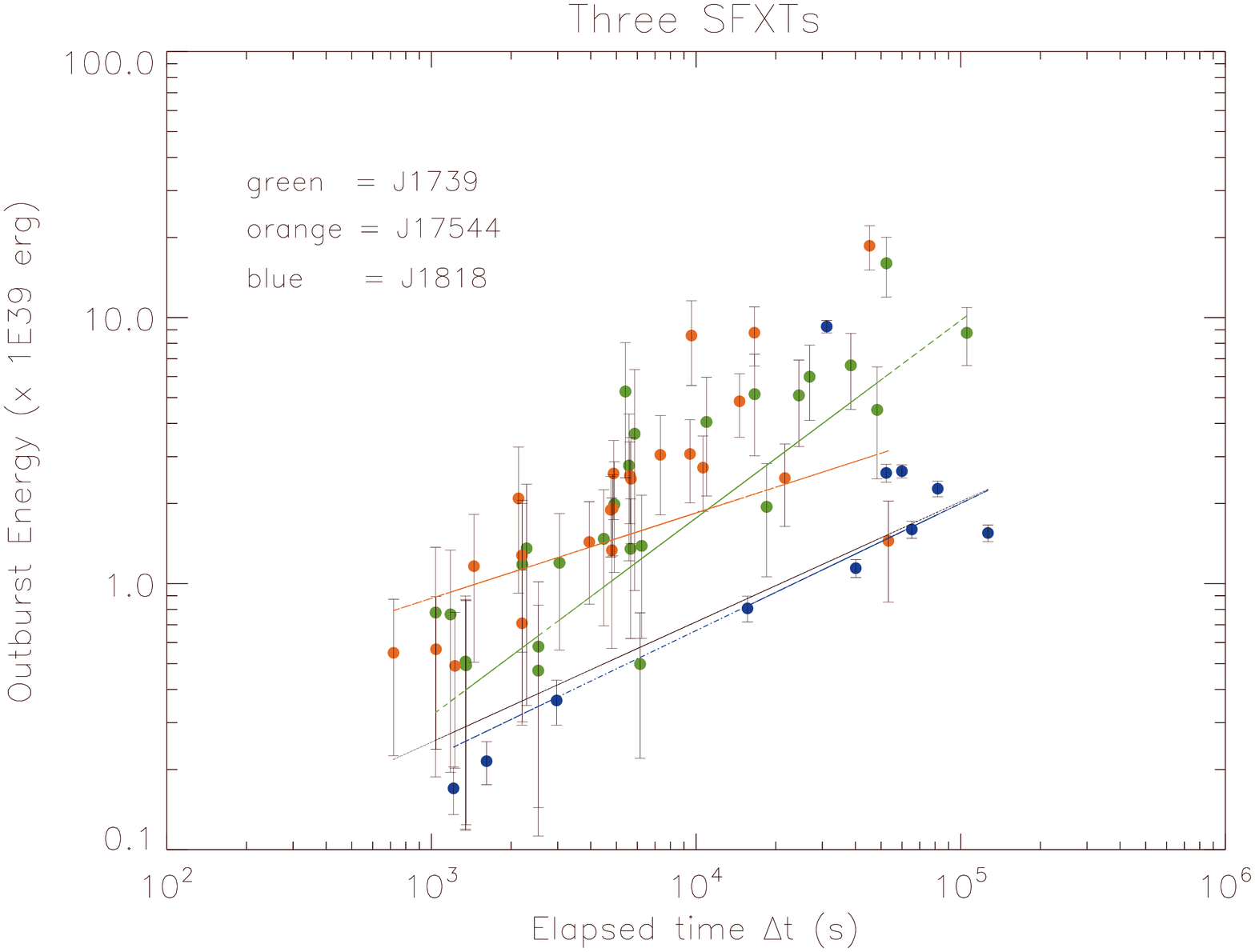} \\
\includegraphics[height=11.cm, angle=0]{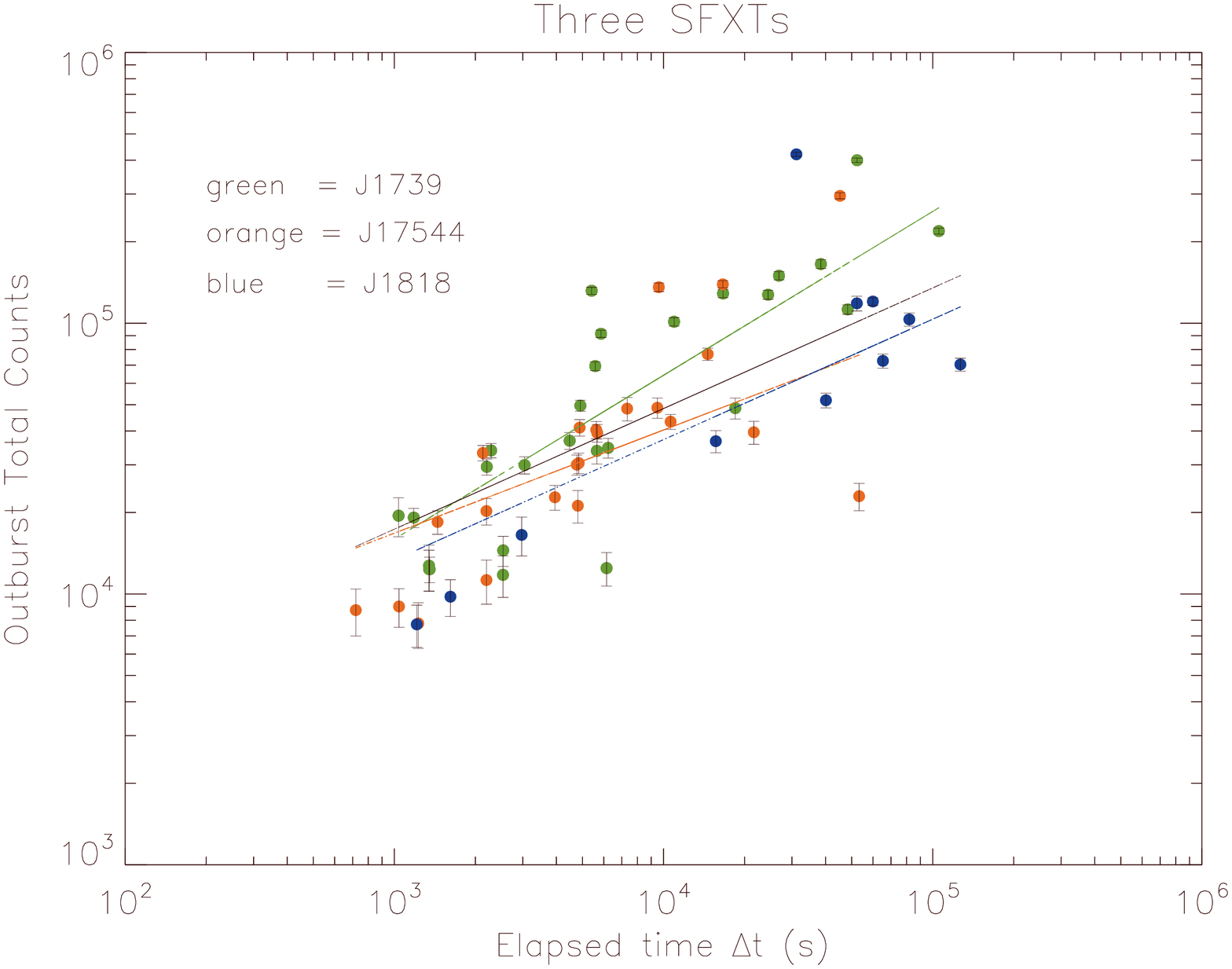}
\end{tabular}
\caption{Summary plots of the power-law fits (in the {\em upper panel} the best-fit power-law 
to the energy versus elapsed times in different colors, 
in the {\em lower panel} the  best-fit power-law to the IBIS/ISGRI counts versus elapsed times. 
The thin black line marks the power-law fit to the data-set of the three sources combined.}
\label{lsfig:allfit}
\end{figure*}

It is seen from Figs.~\ref{lsfig:fit} (summarized also in Figs.~\ref{lsfig:allfit}) 
that there is a correlation between the energy released in the outburst and its duration.
We will discuss this correlation below.

	      %%%%%%%%%%%%%%%%%%%%%%%%%%%%%%%%%%%%%%%%%%%%%%%%%%%%%%%%%%%%%%%%%%%%%
	      \section{Discussion}\label{sec:discussion}
	      %%%%%%%%%%%%%%%%%%%%%%%%%%%%%%%%%%%%%%%%%%%%%%%%%%%%%%%%%%%%%%%%%%%%%

For this study we have used the \inte/IBIS data-set of three SFXTs reported by PS14: \sax, \igr\ and \xte.
We characterized the temporal properties of these flares calculating  the waiting-time between SFXT 
flares, the duration of the outburst phase as observed at hard X--rays
by \inte, and the elapsed times between the first and the last flare belonging to the same oubturst.
We have observed a clustering of bright flares with waiting-times less than $\sim$1~day, 
which was later used to quantify the effective duration of a bright phase of an ``outburst''. 
Finally, we have calculated the cumulative distribution of these time scales (durations and elapsed times) of 
all outbursts from all three SFXTs, finding a power-law-like behaviour.

In PS14 we concentrated on the cumulative distributions of the hard X--ray luminosity of a sample of SFXTs, compared
to other three high mass X--ray binaries (HMXBs), while \citet{Shakura2014} compared distribution of energy of SFXTs flares
with the quasi-spherical settling accretion model, suggesting that the reconnection of magnetic fields carried out by stellar wind from OB-supergiants can be 
the physical mechanism able to trigger the opening of the NS magnetosphere, causing the sudden accretion of the captured matter, and 
producing the X--ray flare.

%%%%%%%%%%%%%%%%%%%%%%%%%%%%%%%%%%%%%%%%%%%%%%%%%%%%%%%%
\begin{figure*}
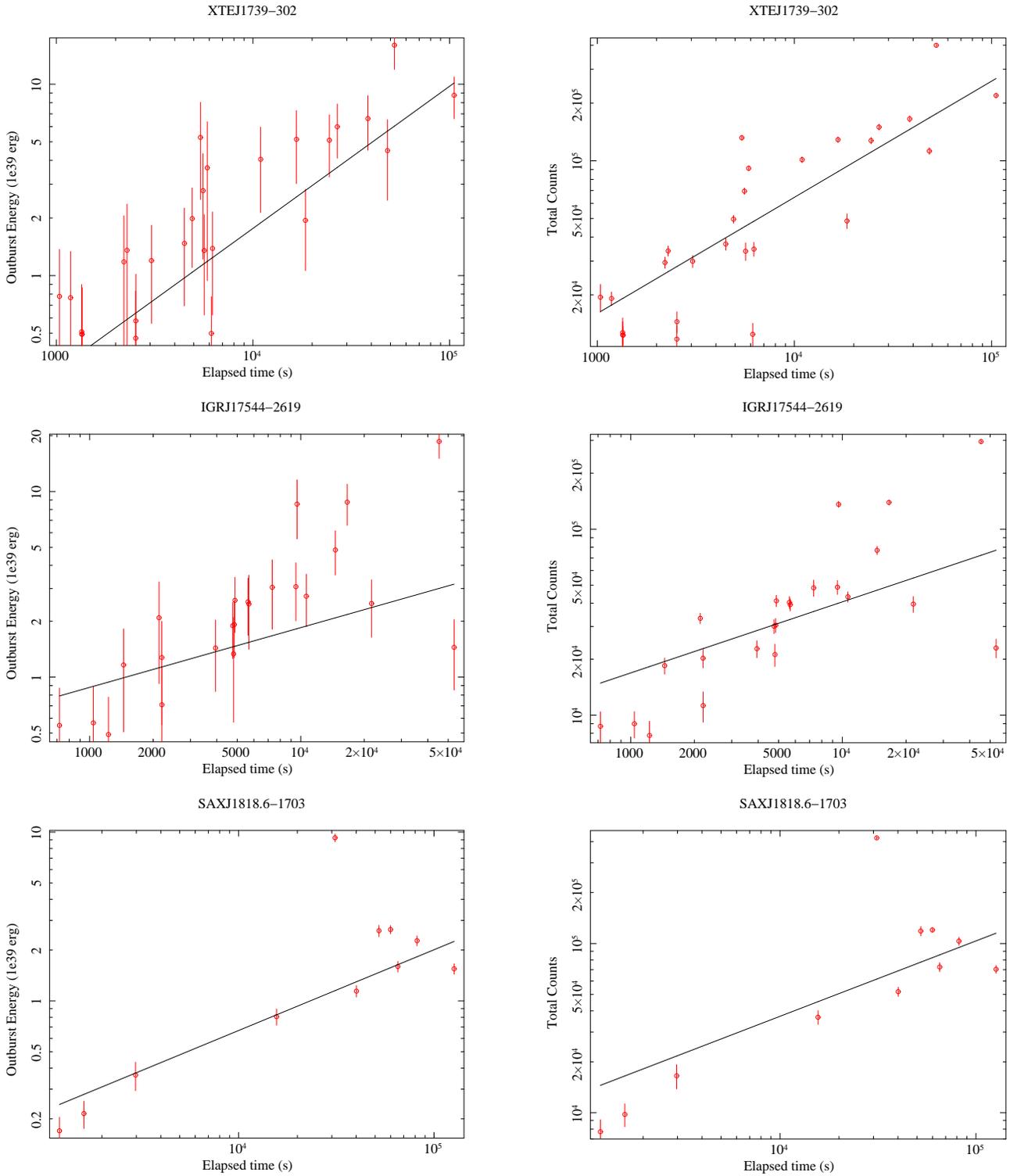

\centering
\begin{tabular}{cc}
\includegraphics[height=8.8cm, angle=-90]{fig6a.ps} & 
\includegraphics[height=8.8cm, angle=-90]{fig6b.ps} \\
\includegraphics[height=8.8cm, angle=-90]{fig6c.ps} & 
\includegraphics[height=8.8cm, angle=-90]{fig6d.ps} \\
\includegraphics[height=8.8cm, angle=-90]{fig6e.ps} & 
\includegraphics[height=8.8cm, angle=-90]{fig6f.ps}  
\end{tabular}
\caption{Power-law fits  to the outburst energy (17--50\,keV) versus elapsed times $\Delta$t (i.e. outburst duration, as defined in the text) of the 3 SFXTs. 
On the {\em left} the panels showing the best-fit power-law to the energy versus elapsed times, on the {\em right} the  best-fit power-law to the  IBIS/ISGRI counts 
versus elapsed times.
Large uncertainties are mainly due to the large uncertainty on the source distances ($\pm{1}$~kpc in XTE~J1739--302  and IGR~J17544--2619; $\pm{0.1}$~kpc in SAX~J1818.6--1703).}
\label{lsfig:fit}
\end{figure*}
%%%%%%%%%%%%%%%%%%%%%%%%%%%%%%%%%%%%%%%%%%%%%%%%%%%%%%%%

%%%%%%%%%%%%%%%%%%%%%%%%%%%%%%%%%%%%%%%%%%%%%%%%%%%%%%%%
%%%%%%%%%%%%%%%%%%%%%%%%%%%%%%%%%%%%%%%%%%%%%%%%%%%%%%%%

%%%%%%%%%%%%%%%%%%%%%%%%%%%%%%%%%%%%
The observed cumulative distributions of temporal properties of the outbursts in three well-studied SFXTs with power-law shapes suggest 
possible self-similar character of the stellar wind properties in these sources,  regardless the different orbital periods. 
It can be shown that such properties naturally arise 
in the frame of the model for bright SFXT flares suggested in \citet{Shakura2014}. We remind that the key feature of the model is the 
settling accretion regime onto a slowly rotating magnetized NS (\citealt{Shakura2012}, \citealt{Shakura2015}). 
This regime can set in if the X-ray luminosity from the NS is below a few times $10^{36}$~erg s$^{-1}$. At this stage, 
a hot convective shell is formed around slowly rotating NS magnetosphere, and the plasma entry to the NS magnetosphere is mediated 
by plasma cooling (Compton at higher accretion rates and radiative at low accretion rates). 
We argued that at the quiescent stage of SFXTs the accretion rate onto the NS is $\dot M_a\simeq \dot M_B f(u)_\mathrm{rad}^{1/3}$, 
where $\dot M_B$ is the standard Bondi-Hoyle-Littleton mass accretion rate 
(determined by the surrounding wind density $\rho_w$ and velocity $v_w$) and $f(u)_\mathrm{rad}\sim 0.03-0.1$ is the reduction 
factor due to radiation plasma cooling.  
A magnetized stellar wind from the optical OB supergiant companion was proposed as trigger for an SFXT flare 
due to reconnection of large-scale magnetic field carried by the wind (Shakura et al. 2014). 
It was shown that the magnetic reconnection preferably occurs at small $f(u)$, which exactly corresponds to low states of SFXTs.
At higher $f(u)$ and, hence, higher plasma entry rate into magnetosphere, the reconnection time is higher than the plasma magnetospheric entry time 
due to instabilities, so the magnetic field is admixed with plasma; this  may cause additional X--ray variability with temporal properties different from 
what is observed, for example, in steady-state accreting X-ray pulsars like Vela X-1 (\citealt{Furst2010}, PS14). 
During each flare initiated by the appearance of open field magnetic lines due to reconnection, the entire mass of the shell around the magnetosphere is accreted onto NS over a time interval corresponding to the free-fall time form the outer boundary of the shell (around the Bondi gravitational capture radius $R_B\approx 2GM/v_w^2$), typically of the order of 1000 s. 

In this picture, the SFXT outburst is represented by a chain of flares which are physically connected to one large region of magnetized stellar wind. Therefore, the temporal properties of flares in the outburst should bear information about the magnetized wind structure. 
Solar wind studies suggest \citep{Zelenyi2004}
that the equatorial magnetized wind agglomerates into a fractal structure with the Hausdorf dimension of $d_f\approx 4/3$, i.e. the mass inside the 
region of size $l$ grows as $M_l \sim l^{d_f}$. 
In the context of the present study, the size of the magnetized stellar wind region is related to the duration of the outburst, $l \sim \Delta t \times v_w$, where $v_w\sim 1000$~km s$^{-1}$ is the typical stellar wind velocity. 
Therefore, Fig.~\ref{fig:all} 
suggests that the longest outbursts $\sim 10^5$ s correspond to the largest size of the magnetized wind clouds of about 100 $R_\odot$, which is commensurable to the orbital separation in these sources.

The gravitational capture (Bondi) radius $R_B$ for the typical wind velocity from O-supergiants $\sim 1000$ km s$^{-1}$ is about 
$2\times 10^{10}$ cm, much smaller than the orbital separation. Therefore, during an outburst of duration $\Delta t$, the volume of the wind captured by NS is $\Delta V_0\sim R_B^2\times v_w\times \Delta t$. 
Suppose this volume to contain $N$ clumps with some mass distribution (which is actually not important for our purposes). 
We stress once again here, while several authors in previous literature have discussed the possibility of direct accretion of wind clumps in SFXTs (e.g. \citealt{Walter2007}), we adopt here the completely different scenario of settling accretion. This means that not only the wind dense clumps, which are present at all times, are responsible for the SFXT activity, but their special properties (e.g., temporarily ejected magnetized lamps of the O-supergiant clumpy wind) are needed to trigger SFXT outbursts.
The mass in the clumps is $\Delta M=\sum_{i=1}^N\rho_i V_i$, where $\rho_i$ and $V_i$ is the density and volume of the $i$-th
clump, respectively. Assume the mean clump density $\rho_1$ for all clumps, which is much larger than the 
density of the surrounding wind. Then the total mass within the volume $V_0$ reads $\Delta M\approx \rho_1\sum_{i=1}^NV_i$, 
neglecting the interclump mass. Therefore, the average density of the wind is $\bar \rho_w=\Delta M/V_0$. 
In the case of the magnetized wind, as we mentioned, during an outburst the magnetospheric instability due to reconnection effectively leads 
to the Bondi accretion regime, i.e. we expect that the total mass accreted onto NS in the entire outburst should be around $\Delta M$. 
As the accreted mass corresponds to the total energy $\Delta E$ released in the outburst, it is expected that 
\begin{equation}
\Delta E\propto \bar\rho_w \Delta t\,.
\end{equation}
If the wind has a fractal structure, i.e. the density in the wind volume 
increases as $\bar\rho_w \propto l^{d_f-3}$, taking into account the relation $l\sim v_w\Delta t$, we find 
\begin{equation}
\Delta E\propto \Delta t^{d_f-2}\,.
\end{equation}
Clearly, in the case of homogeneous (on average, although maybe clumpy) wind density with $d_f=3$ 
the linear dependence of the released energy on the outburst duration, $\Delta E\sim \Delta t$ is expected.
Note that here we assumed a constant wind velocity, which seems to be reasonable in so far as the 
duration of the even longest outburst is much smaller than the orbital period of the binary system.
If we take into account the possible additional wind acceleration between the companion and the NS location, 
the power-law index in 
the above relation $\Delta E \sim \Delta t^b$ will increase: $b> d_f-2$. 

Eq. (2) suggests a pure observational test of the wind structure, which can be 
performed in two ways. 

\begin{figure}
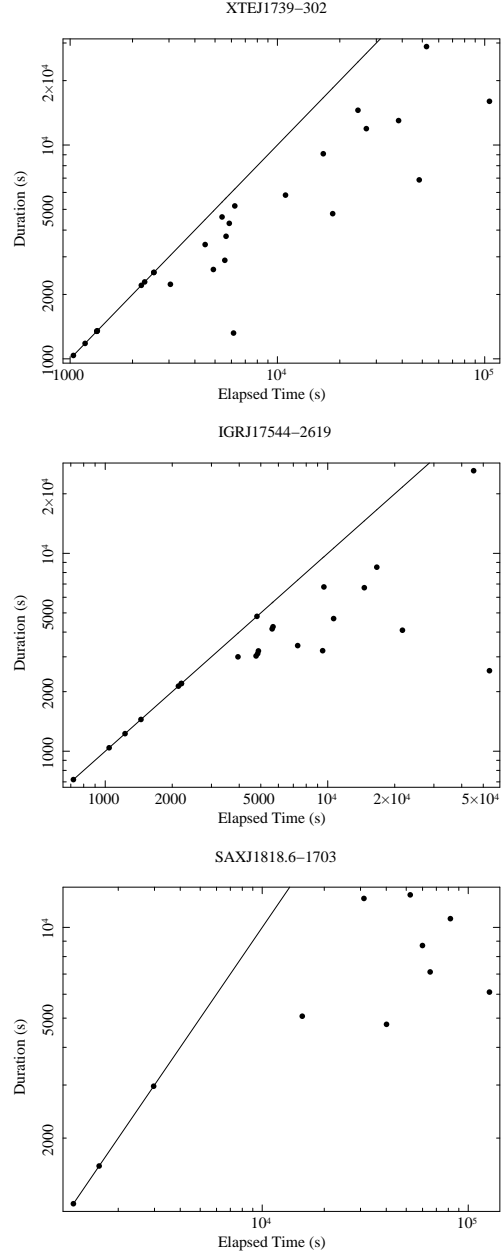

\centering
\begin{tabular}{ccc}
\includegraphics[height=7.3cm, angle=-90]{fig8a.ps} \\
\includegraphics[height=7.3cm, angle=-90]{fig8b.ps} \\
\includegraphics[height=7.3cm, angle=-90]{fig8c.ps}
\end{tabular}
\caption{Total duration of outburst ($D$) versus elapsed times $\Delta$t (i.e. outburst duration, as defined in the text) of the 3 SFXTs. The straight line marks D=$\Delta$t.
}
\label{f:DvDt}
\end{figure}

First, in each particular source, the energy released may be written as 
\begin{equation}
\Delta E = \sum_{i=1}^N \dot M_i \Delta t_i=\langle\dot M\rangle D\,,
\end{equation}
where $\langle \dot M\rangle$ is the mean accretion rate in the flares, which, we remind, 
in the Bondi accretion regime is determined solely 
by the density, $\rho_1$, and velocity, $v_w$, of captured matter (clearly, actually observed variations in
$\dot M_i$ in particular flares reflect variations in density of the corresponding blob being accreted). 
Comparing with Eq. (2) we find the relation 
\begin{equation}
D\propto \Delta t^{d_f-2}\,,
\end{equation}
which includes only observed quantities. Again, in the case of on average homogeneous wind with $d_f=3$,
a linear dependence is expected. In Fig. \ref{f:DvDt} we plot these relation for three SFXTs under study (the straight line marks linearity in the three cases).

As it is unclear how to estimate errors in $D$ and $\Delta t$, it is not possible to perfom a formal fit to the data.
However, we note that in the worst case, on the y-axis the error on the ``D'' values could reach 10\% at most, while on the x-axis 
each elapsed time $\Delta t_i$ is, in the worst case, constrained in the range [$\Delta t_i$, $\Delta t_i$+1~day], by definition.
So, it is clearly seen that non-linearity appears with increasing outburst duration, in all sources. \footnote{The strictly linear behaviour for  $Dt<3000$~s correspond to the 'outbursts' consisting of isolated flares, as mentioned in Section 3, for which $Dt=D$ by definition.}
This is indeed expected, since deviations from homogeneity for (physical) fractal structures are more prominent on large scales.

Obviously, increase in the outburst statistics would allow a more precise statistical analysis.

%%%%%%%%%%%%%%%%%%%%%%%%%%%%%%%%%%%%%%%%%%%%%%%%%%%%%%%%
%%%%%%%%%%%%%%%%%%%%%%%%%%%%%%%%%%%%%%%%%%%%%%%%%%%%%%%%

The second way to test Eq.(2) is to directly compare the energy released in each outburst with its duration. 

Comparison with the observed correlation (see Fig.~\ref{lsfig:fit} and Table~\ref{lstab:fit}) shows that, on average, $b \approx 0.5$, 
implying $d_f-2<0.5$ and hence $d_f<2.5$ for stellar winds in the studied SFXTs. 
This suggests a fractal structure of the OB-supergiant winds accreting onto NS in these sources. 
Of course, the radiatively-driven winds from OB-supergiants should have different properties than the solar wind, 
but the appearance of fractal structures in magnetized plasma flows seems very plausible. 

In the proposed picture, the matter is gravitationally captured by the NS and 
is stored in the shell around the magnetosphere until the reconnection of the captured magnetic field occurs. 
The duration of refilling the magnetospheric shell by gravitationally captured matter should occur on the same time scale 
as the flare itself (of the order of the free-fall time from the Bondi radius, a thousand of seconds), 
which leads to a continuous sequence of flares during the entire time of 
the outburst due to accretion of magnetized clump of stellar wind.  
The observed nearly power-law distribution of durations of SFXT outbursts reflects the size distribution 
of magnetized clouds in the wind of the OB-supergiant companions.

Clearly, an increased  statistics of the flares and outbursts from SFTXs as well as spectroscopic observations in other bands are needed 
to understand properties of the magnetized winds from OB-supergiants more in depth.

%%%%%%%%%%%%%%%%%%%%%%%%%%%%%%% 
\begin{table}
 \centering
  \caption{Results obtained fitting the outburst Energy (or total IBIS/ISGRI counts in the energy band 17--50 keV) versus 
the outburst duration (``Elapsed time'') with a power-law model, defined as y = ax$^b$. Quoted uncertainties are at 1$\sigma$.}
  \begin{tabular}{@{}llll@{}}
\hline
Source                    &    norm {\em a}    &    pow slope  {\em b}   &   $\chi ^2$ (dof) \\
\hline
Counts vs Elapsed time &          &          &         \\
\hline
XTE~J1739--302          &    244 $\pm{17}$                   &    0.606  $\pm{0.007}$       &    2912 (25)   \\
IGR~J17544--2619        &    1210 $^{+140} _{-110}$          &    0.38  $\pm{0.01}$         &    2386 (21)    \\
SAX~J1818.6--1703       &    620   $\pm{80}$                 &    0.44 $^{+0.02} _{-0.01}$  &    615.1 (9)    \\
All sources             &    800   $\pm{40}$                 &    0.446 $\pm{0.004}$        &    9637 (59)     \\
\hline
Energy vs Elapsed time &          &          &         \\
\hline
XTE~J1739--302          &    $0.0019 ^{+0.0020} _{-0.0010}$   &     0.74 $\pm{0.08}$      &   29.40  (25)   \\
IGR~J17544--2619        &    $0.10 ^{+0.06} _{-0.04}$         &     0.32  $\pm{0.06}$     &   60.70  (21)   \\
SAX~J1818.6--1703       &    $0.0082 ^{+0.0019} _{-0.0016}$   &     0.48  $\pm{0.02}$     &  403.0    (9)   \\
All sources             &    $0.011 \pm{0.002}$      &     0.45  $\pm{0.02}$     &    582 (59)    \\
\hline
\end{tabular}
\label{lstab:fit}
\end{table}
%%%%%%%%%%%%%%%%%%%%%%%%%%%%%

Finally, it is interesting to note that the behaviour of the three prototypical SFXTs discussed here during the bright phase of their outbursts is very similar, irrespective of the wide range of 
orbital periods covered ($\sim$5, 30 and 51 days; PS14, \citealt{Walter2015}), likely due to both the intrinsic wind properties and to the mediating role of the 
shell above the NS magnetosphere, in the settling accretion scenario.

%%%%%%%%%%%%%%%%%%%%%%%%%%%%%%%%%%%%%%%%%%%%%%%%%%%%%%%%%
\section*{Conclusions}
%%%%%%%%%%%%%%%%%%%%%%%%%%%%%%%%%%%%%%%%%%%%%%%%%%%%%%%%%

We have performed a characterization of the temporal properties of X--ray flares from the three most extreme
SFXTs, as observed by \inte\ in the energy band 17--50 keV,
obtaining more insights into the physics producing their outbursts.

Calculating the cumulative distribution of waiting-times between the X--ray flares, 
we were able to identify the typical timescale that clearly separates different outbursts, 
\text{each} composed by several single flares at $\sim$ks timescale.
This selection allowed us to calculate the energy emitted during SFXT outbursts, finding an interesting correlation
with the outburst duration.

In the framework of the quasi-spherical settling accretion we have discussed here, 
the outburst properties (total emitted energy, duration and their positive correlation) 
carry signatures of the magnetized wind structure of the companion:
an SFXT outburst is composed by a chain of
X--ray flares physically connected to one large region of magnetized
stellar wind that triggers the NS magnetospheric instability by means of magnetic reconnection.

The power-law slope of the correlation between total emitted energy and duration of the SFXT outbursts 
can be explained by the fractal structure of the OB-supergiant winds, reflecting the 
size distribution of magnetized clouds in the wind of the OB-supergiant companions.

%%%%%%%%%%%%%%%%%%%%%%%%%%%%%%%%%%%%%%%%%%%%%%%%%%%%%%%%%
\section*{Acknowledgments}
%%%%%%%%%%%%%%%%%%%%%%%%%%%%%%%%%%%%%%%%%%%%%%%%%%%%%%%%%

Based on observations with \textit{INTEGRAL}, an ESA project
with instruments and science data centre funded by ESA member states
(especially the PI countries: Denmark, France, Germany, Italy,
Spain, and Switzerland), Czech Republic and Poland, and with the
participation of Russia and the USA. 
This work has made use of the \inte~archive developed at INAF-IASF Milano, 
http://www.iasf-milano.inaf.it/$\sim$ada/GOLIA.html.
LS and AP acknowledge the Italian Space Agency financial support INTEGRAL
ASI/INAF agreement n. 2013-025.R.0, and the grant from PRIN-INAF 2014, 
``Towards a unified picture of accretion in High Mass X-Ray Binaries'' (PI: Sidoli).
The work of KP is partially supported by RFBR grant 14-02-00657.
We thank the anonymous referee whose suggestions helped improve and clarify the
manuscript.
\bibliographystyle{mn2e} 
\bibliographystyle{mnras}
%\bibliography{biblio}

\bsp

\label{lastpage}

\end{document}